\documentclass[useAMS,usenatbib, usegraphicx]{mn2e}

\usepackage{amsmath}
\usepackage{amssymb}
\title[Dark matter and dark energy accretion onto IMBHs]{Dark matter and dark energy
accretion onto intermediate-mass black holes}
\author[C. Pepe et al.]{C. Pepe$^{1,2}$\thanks{E-mail:
carolinap@iafe.uba.ar}, L. J. Pellizza$^{1,2}$ and G. E. Romero$^{2,3}$\\
$^{1}$Instituto de Astronom\'ia y F\'isica del Espacio, Casilla de Correo 67,
Suc. 28, 1428, Buenos Aires, Argentina\\
$^{2}$Consejo Nacional de Investigaciones Cient\'ificas y T\'ecnicas, CONICET,
Argentina\\
$^{3}$Instituto Argentino de Radioastronom\'{\i}a, Argentina}
\begin{document}

\date{Accepted 2011 November 23. Received 2011 November 18; in original form 2011 October 12}

%\pagerange{\pageref{firstpage}--\pageref{lastpage}} \pubyear{2011}
\pagerange{3298--3302} \pubyear{2012}
\volume{420}

\maketitle

\label{firstpage}

\begin{abstract}
In this work we investigate the accretion of cosmological fluids onto an
intermediate-mass black hole at the centre of a globular cluster, focusing on
the influence of the parent stellar system on the accretion flow. 
%We integrate
% the general-relativistic hydrodynamical equations for a fluid in the presence
% of the gravitational field of the cluster and the black hole, and compute the
% accretion rate onto the compact object. 
We show that the accretion of
cosmic background radiation and the so-called dark energy onto an intermediate-mass
black hole is negligible. On the other hand, if cold dark matter has a
nonvanishing pressure, the accretion of dark matter is
large enough to increase the black hole mass well beyond the present observed
upper limits. We conclude that either intermediate-mass black holes do not
exist, or dark matter does not exist, or it is not strictly collisionless. In the
latter case, we
set a lower limit for the parameter of the cold dark matter equation of state.
\end{abstract}


\section{Introduction}

Intermediate-mass black holes (IMBHs) are hypothetical compact objects with
masses in the range $10^2$--$10^4\,M_\odot$, proposed to explain ultraluminous 
X-ray sources (ULXs), which exceed by 1--2 orders of magnitude the
Eddington limit for the accretion luminosity of stellar-mass black holes
\citep{angelini,fabbiano}. These objects are thought to form in the centers
of globular clusters, where the stellar density is high enough to trigger the
runaway merging of stellar-mass compact remnants
\citep{portegies,gurkan,freitag}. Careful measurements of stellar kinematics
and density profiles  of globular clusters suggest indeed that some of these
clusters may harbour IMBHs, although the evidence is still not conclusive
\citep[e.g.][]{noyola, baumgardt}. The properties of an extended central X-ray source in NGC~6388
also suggest an IMBH as a possible explanation for its central engine
\citep{nucita}. The fact that many ULXs in nearby galaxies are associated with
globular clusters provides another argument for their existence. However,
alternative explanations can be found for these observations, hence the
existence of IMBHs is still a matter of discussion.

The detectability of IMBHs depends on their masses, since the effects on their
surroundings increase with mass. The mass of an IMBH is
determined by its initial mass and the accretion of matter and energy from their surroundings.
This accretion can occur in discrete events, such as the infall of the
stripped envelope of a star passing near the IMBH \citep[e.g.][]{miocchi}, or
by the continuous infall of any media in which the IMBH is immersed, such as
cosmological fluids (dark energy,
dark matter, the cosmic microwave background) or the intracluster medium of the
globular cluster.

The solution of the steady spherical accretion flow of a classical fluid onto an
isolated compact object has been obtained by \citet{bondi}. \citet{michel}
worked out the corresponding solution of the general relativistic fluid
equations in the Schwarzchild metric, considering a fluid composed of massive
particles, while \citet{babichev} developed the same model for massless
particle fluids. In all cases, an important prediction of the models is that
the accretion rate scales as the square of the mass of the compact object.
Applying \citet{babichev} model to an IMBH, one can readily see that the
total mass change due to the accretion of dark energy during the entire life
of the IMBH is negligible. However, this might not be the case for dark matter.
Several authors have investigated the accretion of dark matter
by supermassive black holes \citep{macmillan, zelnikov, zelnikov2, munyaneza, peirani}. 
 \citet{guzman} computed time-dependent accretion models, finding that if dark matter is
pressureless, it must contribute with a major fraction of the present black hole
mass. 
The same reasoning might be applied to IMBHs, suggesting that they
might have accreted hundreds to thousands of solar masses in their lifetime.

However, IMBHs are not isolated objects; they reside in the center of
massive stellar systems. The gravitational field of these systems might influence
the accretion flow, enhancing the accretion rate, an effect not taken into
account in the models of \citet{babichev} and \citet{guzman}. Hence, an
investigation of the effects of the host stellar systems of IMBHs on the
accretion flow seems relevant, as it would allow to make more precise
predictions about the accretion rates onto IMBHs and their final masses, which
can be compared with the available observations.

In this paper we explore simple models for the steady spherical accretion flow
onto an IMBH located at the center of a globular cluster, taking into account
the influence of the cluster. In Section~\ref{model} we develop our models,
based on relativistic fluid dynamics in the gravitational field of both the
black hole and the cluster. In Section~\ref{results} we apply them to
compute the accretion rates for different kinds of fluids and we estimate the final
masses of IMBHs. Finally, in Section~\ref{conclusions} we discuss the impact of
our results on the current knowledge of IMBHs and summarize the results of our work.

\section{The model}
\label{model}

Our model is based on the following hypotheses. The accretor is an IMBH of mass
$M$ located at the center of a globular cluster. The mass distribution of the
stellar system (cluster plus black hole) is assumed to be spherical, hence the
space-time interval can be written as $ds^2 = e^\nu dt^2 - e^\lambda dr^2 -
r^2 (d\theta^2 + \sin^2\theta d\phi^2)$, where $r,\theta,\phi$ are the usual
spherical coordinates, $t$ is the time and $\nu$ and $\lambda$ are functions of
$r$ only. These functions are determined by the mass distribution of the
stellar system, which in turn can be described by a single function $m(r)$ that
gives the mass $m$ enclosed by a sphere of radius $r$. The Schwarzschild metric
is the particular case in which $m(r) = M$, and hence $\nu = -\lambda =
\ln(1 - 2 M / r)$. Throughout this work, we use the natural system of units in
which the gravitational constant $G$ and the speed of light $c$ are equal to
unity. The accreted medium can be described as a relativistic fluid with an
equation of state $p = \omega \rho$, where $p$ is its pressure and $\rho$ its
energy density. Its energy-momentum tensor is $T_{\mu\nu} = (\rho + p) u_\mu u_\nu
- p g_{\mu \nu}$, where $g_{\mu\nu}$ is the metric tensor and $u^\mu = dx^\mu{/ds}$
is the fluid four-velocity with $u^\mu u_\mu = 1$.

We aim at computing the accretion rate onto the IMBH, $\dot{M} = -4 \pi
\lim_{r \to 2 M} r^2 T^r_0$, for which we integrate the equations of motion of
the fluid, subject to the boundary conditions that far from the globular
cluster the density $\rho_\infty$ and pressure $p_\infty$ are constant, and the
fluid is at rest ($u_\infty = 0$). Following \citet{babichev},
the integration of the time component of the energy-momentum conservation law
$T^{\mu \nu}_{;\nu} = 0$ gives the energy conservation equation,

\begin{equation}
 (p + \rho) \left(e^{-\nu} + e^{\lambda -\nu} u^2\right)^{1/2} u r^2 e^{\frac{1}{2}
(\lambda + 3 \nu)} = C_1,
\label{bernoulli}
\end{equation}

\noindent where $C_1$ is a constant of integration.
Another integral of motion is obtained by projecting the energy-momentum
conservation law on the four velocity, $u_\mu T^{\mu \nu}_{;\nu} = 0$. Integrating
this equation gives the energy flux equation 

\begin{equation}
 e^{\int_\infty^\rho \frac{d\rho'}{\rho' + p}} e^{\frac{1}{2} (\nu + \lambda)} u r^2 = C_2,
\label{energia}
\end{equation}

\noindent where $C_2$ is another constant. Combining eqns.~\ref{bernoulli} and
\ref{energia} we obtain

\begin{equation}
(p + \rho) e^{-\int_\infty^\rho \frac{d\rho'}{p + \rho'}} \left(1 +
e^\lambda u^2\right)^{1/2} = \rho_\infty + p_\infty.
\label{integral_critico}
\end{equation}

\noindent
Considering the expression shown above, the accretion rate can be evaluated as

\begin{equation}
 \dot{M} = -4 \pi (\rho_\infty + p_\infty) C_2, 
\label{tasa}
\end{equation}

\noindent where we have used that in the limit $r \to 2 M$ the influence of the
globular cluster vanishes and the metric approaches to that of Schwarzschild,
for which $\lambda + \nu = 0$.

The constant $C_2$ determines the accretion flux
onto the black hole and can be calculated by fixing the parameters at any
point. \citet{michel} has shown that the relativistic hydrodynamical
equations for accretion onto an isolated black hole have a critical point,
which can be used to determine $C_2$.
To obtain the critical point in our case we follow \citet{michel} and write
eqns.~\ref{bernoulli} and \ref{energia} in their differential form,
combining them into

\begin{eqnarray}
 \left[ (e^{-\nu}+e^{\lambda -\nu} u^2)^{-1} e^{\lambda-\nu}u - \frac{\omega}{u} \right] du + \nonumber \\
\left[ \frac{1}{2} (e^{-\nu}
 + e^{\lambda - \nu} u^2)^{-1} (-e^{-\nu} \nu'+ (\lambda'- \nu')u^2 e^{\lambda - \nu}) +
\right.
\nonumber \\
\left. \frac{\lambda'}{2} + \frac{3\nu'}{2} + \frac{2}{r} - (1 + \omega) \left(\frac{\nu'}{2} + \frac{\lambda'}{2} + \frac{2}{r}
 \right) \right] dr = 0,
\label{flowdif}
\end{eqnarray}

\noindent where we used the equation of state. We can see
from this expression that if one of the bracketted factors vanishes, there is
a turn-around point and the solutions are double-valued. Only solutions that
pass through the critical point where both brackets vanish simultaneously are
single valued, leading to 

\begin{equation}
 \frac{1}{2} \nu'(r_{c})(1 - \omega) - \frac{2 \omega}{r_{c}} = 0, 
\label{critico}
\end{equation}

\noindent and

\begin{equation}
 u_c^2 =  \frac{\omega}{1 - \omega}. 
\label{ucritico}
\end{equation}

\noindent Here, the subscript ``c'' stands for the corresponding variables
evaluated at the critical values. It is apparent from eqn.~\ref{ucritico} that
there is no critical point for $\omega < 0$ (dark energy). We will treat this
case separately later, and focus now on the case $\omega > 0$. \citet{misner}
have shown that for a spherical stellar system,

\begin{equation}
\nu'(r)= \frac{2 m(r)}{r \left[ r - 2 m(r)\right]}.
\end{equation}

\noindent Hence $r_c$ is defined by the implicit equation

\begin{equation}
\frac{m(r_c)}{r_c} = \frac{2 \omega}{1 + 3 \omega}.
\label{rcritico}
\end{equation}

\noindent
Evaluating eqns.~\ref{energia} and \ref{integral_critico} at the critical
point, we replace $C_2$ in eqn.~\ref{tasa} to obtain the final expression in
terms of known quantities,

\begin{eqnarray}
\dot{M} = -\pi (1 + \omega) \rho_{\infty}\left[ e^{\nu_c}
(1 + \frac{\omega}{1-\omega})\right]^{-\frac{1}{2\omega}} \nonumber \\
\left(\frac{\omega}{1-\omega} \right)^{\frac{1}{2}} m^2(r_c)
\left(\frac{1+3\omega}{\omega}\right)^2 e^{\frac{1}{2}\nu_{c}}.
\label{acrecion_final}
\end{eqnarray}

\noindent
Note that $\dot{M}$ is proportional to $m^2(r_{\mathrm{c}})$ which, depending
on the location of the critical point, can range from $M^2$ to $(M_{\mathrm{gc}} + M)^2$,
where $M_{\mathrm{gc}}$ is the mass of the globular cluster. Hence, as $M_{\mathrm{gc}} \gg M$, the accretion can be
greatly enhanced if the critical point is located in the outer regions of the
cluster. Eqn.~\ref{rcritico} shows that for a given cluster, the location of
the critical point depends solely on $\omega$. 

In the next sections we apply
this model to several fluids with different values for the equation of state
parameter $\omega$, and analize the resulting accretion rates.

\section{Accretion rates}
\label{results}

\subsection{Perfect relativistic fluid}
\label{perfect_fluid}

\begin{table}
\begin{center}
\caption{Model parameters.}
\begin{tabular}{l c }
%\hline

\hline
Cluster scale radius ($r_0$) & 0.35 pc  \\
Cluster tidal radius ($r_t$) & 44 pc  \\
Cluster concentration ($c_{\mathrm{gc}}$) & 1.8 \\
Cluster galactocentric distance ($R_{\mathrm{gc}}$) & 3.1 kpc \\
Black hole mass ($M$)  & $3000\, M_{\odot}$ \\
Dark matter density at $R_{\mathrm{gc}}$ ($\rho^{\mathrm{DM}}$) & $4.0\times 10^{-21} \mathrm{kg\ m}^{-3}$ \\
Dark energy density ($\rho^{\mathrm{DE}}$) & $7.7 \times 10^{-27} \mathrm{kg\ m}^{-3}$\\
\hline
\label{tabla}
\end{tabular}
\end{center}
\end{table}

As an example we consider the perfect relativistic fluid, for which
$\omega$ is of the order of unity and equal to the square of the sound speed
in the medium in units of $c$. This model describes, for example, the cosmic microwave
background ($p = \rho / 3$). The description of the globular cluster plus
black hole mass distribution $m(r)$ was taken from the model of \citet{miocchi},
with a scale radius $r_0 = 0.35~{\rm pc}$, a concentration $c_{\mathrm{gc}} = 1.8$, and a central
black hole mass of $M = 3000 M_{\odot}$. These parameters, taken from \citet{harris}, were chosen to
represent the mass distribution of NGC~6388, one of the globular clusters with the
strongest evidence in favour of the existence of an IMBH.

\begin{figure}
 \centering
 \includegraphics[width=0.75\columnwidth]{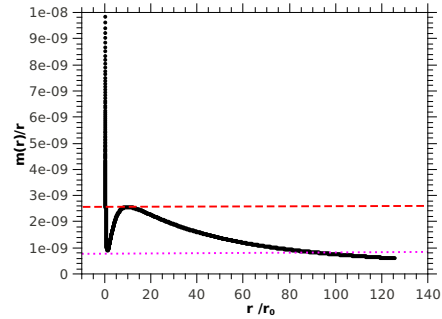}
 \caption{Ratio $m(r)/r$ for a typical globular cluster hosting an
IMBH. The dashed and dotted lines define respectively the maximum and minimum
values for
$2 \omega / (1 + 3 \omega)$ for which eqn.~\ref{rcritico} has three solutions.}
 \label{fig:modelo}
\end{figure}

Eqn.~\ref{rcritico} shows the individual
contribution of the black hole and the globular cluster in a very practical way.
A plot of the left-hand-side of this  equation for our model is shown in Fig. \ref{fig:modelo}. The globular
cluster contribution is the small peak at $r/r_0 \sim 10$, while that of the
IMBH is the asymptotic growth at $r \to 0$. We can see that for relativistic
fluids ($\omega \sim 1$) the right-hand-side of eqn.~\ref{rcritico} takes values of the
order of unity, several orders of magnitude higher than the maximum globular
cluster contribution. The critical radius then occurs at $r \ll r_0$, where the
black hole dominates the scene, and we can approximate the metric by that of
Schwarzschild and solve for $r_{\mathrm{c}}$, $u_{\mathrm{c}}$ and $\dot{M}$. The problem in this case
reduces to that of an isolated IMBH, and the accretion
rate depends only on $M^2$, as predicted by \citet{babichev}. 
As a toy example, for our model, the accretion rate of the
cosmic microwave background is
$\dot{M} = 1.7 \times 10^{-29}\,M_{\odot}\,{\rm yr}^{-1}$,
a completely negligible value resulting from both the $M$-dependence of
$\dot{M}$ and
the extremely low photon density of the cosmic background during the
matter-dominated era in which stellar systems formed. This value is in excellent agreement with the results of \cite{freitas}.

\subsection{Cold dark matter}
\label{DM}

 Cold dark matter is a non-relativistic fluid with negligible thermal
energy and pressure, hence $\rho$ is essentially the rest energy density. For
this fluid, the equation of state $p = \omega \rho$ with $\omega \ll 1$ can be used if we assume an
isothermal flow. The same description applies to any perfect baryonic
fluid at low temperature. Fig.~\ref{fig:modelo} shows that for a fluid
with $\omega \lesssim 4 \times 10^{-8}$, the
accretion flow is indeed modified by the presence of the globular cluster.

\begin{figure}
\centering
\includegraphics[width = 0.75\columnwidth]{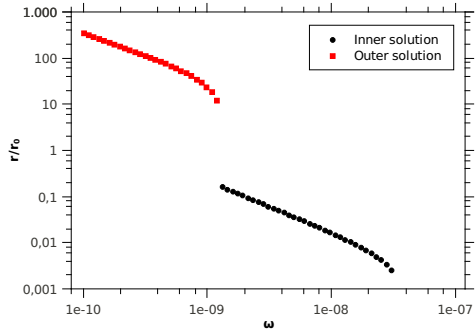}
\caption{Critical radius for a typical globular cluster/IMBH system, as a
function of the equation of state parameter $\omega$. Circles
and rectangles correspond to the inner and outer solutions of
eqn.~\ref{rcritico}, respectively.}
\label{fig:radios}
\end{figure}

%This is the case of cold dark matter, which is assumed to have a very low or
%negligible pressure, or any perfect baryonic fluid at temperatures
% $T \lesssim 10^3\,{\rm K}$. 
%We can see in Fig.~\ref{fig:modelo} that this case
%is cualitatively different from the scenario presented in section
%\ref{perfect_fluid}. 
For very low values of $\omega$ ($\omega \lesssim 10^{-9}$, 
below the dotted line in Fig.~\ref{fig:modelo}), the
critical point lies far from the cluster center (Fig.~\ref{fig:radios}),
and $m(r_{\mathrm{c}}) \approx M + M_{\mathrm{gc}}$. Hence, we expect the accretion rate to be
greatly enhanced. There is also a transition region in which
eqn.~\ref{rcritico} has three solutions (between the dotted and dashed lines
in Fig.~\ref{fig:modelo}). However, only two of them are physically admissible.
If the middle one were the critical radius, the other two must be turn around
points, violating the boundary conditions on the flow velocity. Hence, only the
inner and outer solutions are possible, but only one of them corresponds to
the critical radius. This radius can be found by integrating the flow
equation~\ref{flowdif} for both solutions. Only in one case the integration
converges, indicating that this is the correct value of $r_{\mathrm{c}}$. In the case
of the cluster considered here, the full integration of the flow shows that the critical
point is the inner solution for $\omega > \omega_t = 1.5 \times 10^{-9}$, and the
outer one otherwise. At this transition value, the position of the critical
point changes abruptly (Fig.~\ref{fig:radios}). This change, probably an artifact due to the
simplicity of our model, does not affect our conclusions as we will show later.

\begin{figure}
\centering
\includegraphics[width = 0.8\columnwidth]{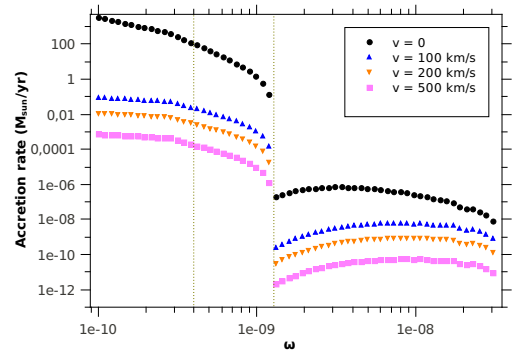}
\caption{Accretion rate for a typical globular cluster/IMBH system, as a
function of the equation of state parameter $\omega$, for different velocities
of the cluster relative to the fluid. Dotted-vertical lines correspond to
the upper and lower limits of the transition region.}
\label{fig:tasas}
\end{figure}

In Fig.~\ref{fig:tasas} we show the results for the accretion rates derived
from eqn.~\ref{acrecion_final}, and using $\rho_\infty = \rho^{\mathrm{DM}} c^2$ (see Table~\ref{tabla}),
where $\rho^{\mathrm{DM}}$ is the dark matter density at the galactocentric distance of the globular
cluster, taken from the work of \citet{klypin}.
We can see the abrupt change in the accretion rate, which
follows the corresponding change in $r_{\mathrm{c}}$. 
%For $\omega > \omega_t$ the accretion rate
%is negligible ($\dot{M} < 10^{-12}\,M_{\odot}\,{\rm yr}^{-1}$), while for $\omega >
%\omega_t$ it grows abruptly following the corresponding change in $r_{\mathrm{c}}$. 
At
$\omega_u \approx 3 \times 10^{-10}$, $r_{\mathrm{c}}$ reaches the globular cluster tidal
radius. For $r \gg r_t$ the globular cluster cannot be considered an
isolated object any more, as the tidal field of the Galaxy becomes dominant.
The predictions of our model are not reliable in such a case. 
% However, in such region, between
%$\omega_{\mathrm{t}}$ and $\omega_{\mathrm{u}}$, the accretion rate is in the range
%$10^{-5}$--$10^{-3}\,M_{\odot}\,{\rm yr}^{-1}$. 

It is important to discuss the effect of the cluster (and thus the IMBH)
motion relative to the dark matter fluid on the accretion rate previously
calculated. \citet{hoyle} have derived a formula for the accretion rate of a
classical fluid by an object moving through it, which is similar to that of
\citet{bondi}. According to the formula of \citet{hoyle}, the relation between
the accretion rate $\dot{M}(v)$ onto an object moving with velocity $v$ with
respect to the fluid, and the accretion rate $\dot{M}_0$ onto an object at rest
is 

\begin{equation}
\label{corrvel}
\dot{M}(v) = \dot{M}_0 \frac{c_{\mathrm{s}}^3}{\left(v^2 + c_{\mathrm{s}}^2 \right)^{3/2}},
\end{equation}

\noindent 
where  $c_{\mathrm{s}}^2 = c^2 \omega$ is the sound speed in the medium. 
Computing the Newtonian limit for eqn.~\ref{acrecion_final}, it can be easily
seen that in the limit $\omega \rightarrow 0$ the accretion rate approaches
that of \citet{bondi}. Hence, we assume that the correction factor given by
eqn.~\ref{corrvel} can be used to derive $\dot{M}(v)$ from $\dot{M}_0$ in
our models. In Fig. \ref{fig:tasas} we show the accretion 
rate values corresponding to cluster velocities $v$ equal to 0, 100, 200 and 500 km/s, which are typical for the motion of Galactic halo objects such as
globular clusters. 

If we neglect the cosmological evolution of the dark matter density, we can
compute a rough estimate of the amount of matter $\Delta M$ accreted by the black hole
along the globular cluster lifetime $\Delta t \sim 10\,{\rm Gyr}$. For $\omega \lesssim 10^{-9}$, $\dot{M}$
depends on $m(r_{\mathrm c}$, with $r_{\mathrm c} \gg r_0$, hence the change in the accretion rate 
due to the IMBH growth is negligible. Hence, we can compute $\Delta M \approx \dot{M} \Delta t \gtrsim
10^4\,M_\odot$ for any reasonable value of $v$. This is not consistent with present upper limits of
\citet{lanzoni,lutzgendorf,cseh} for the mass of the IMBH in NGC 6388.  On the other hand, for $\omega \gtrsim 10^{-9}$ the 
accretion rate is low enough to avoid this discrepacy. This result is remarkable, as it implies that
either the globular cluster does not host an IMBH, or there is a stringent lower limit
on the cold dark matter equation of state parameter, which cannot be null but
at least of the order of $10^{-9}$.

\subsection{Dark energy}
\label{DE}

In previous sections, we based our calculations on the existence of a critical
radius and, hence, a critical velocity for the flow. However, we can see from
eqn.~\ref{rcritico} that for unstable fluids ($\omega < 0$; see 
\citealt{carroll,fabris} for reviews on this subject)
 there is no
critical point outside the blackhole horizon ($r_{\mathrm{c}} < 2 M$). 
% It is important to
% remind what this critical point means: an accreting flux has such a point if
% its velocity increases from subsonic to supersonic values. In an unstable
% fluid ($c_{\mathrm{s}}^2 < 0$; see \citealt{carroll} and \citealt{fabris} for reviews on 
% the subject) the velocity never crosses such a point. 
Following
\citet{michel} and \citet{babichev}, we assume that in this case the
instabilities in the flow cause the growth of the fluid velocity up to the
speed of light at the black hole horizon, and use this as a boundary
condition instead of the critical values. With this assumption, the accretion
rate of dark energy becomes identical to that of \citet{babichev},

\begin{equation}
 \dot{M} = 16 \pi (1 + \omega) \rho_{\infty} M^2.
\label{mpunto_de}
\end{equation}

\noindent
This has a simple interpretation: in this regime, the pressure forces of a
relativistic fluid are of the order of the gavitational forces produced by
the black hole near its horizon, hence much greater than those of the
gravitational field of the cluster elsewhere, and the system behaves as if the
black hole were isolated, as in the model of \citet{babichev}. Another
interesting
property of eqn.~\ref{mpunto_de} is that $\dot{M}$ is negative when
$\omega < -1$ (i.e., for phantom energy), leading to a decrease of the
black hole mass. 
It is important to point out that this would be the
case only if the Generalized Second Law is violated, as \citet{horvath} 
have shown.
For our cluster
plus black hole model, and using the present dark energy density $\rho^{\mathrm{DE}} c^2$ as $\rho_\infty$ (see Table~\ref{tabla}), we obtain

\begin{equation}
\dot{M} = 9.5 \times 10^{-34} \,M_{\odot}
\,{\rm yr}^{-1} (1 + \omega) \left(\frac{M}{M_{\odot}}\right)^2.
\label{mpunto_de_val}
\end{equation}

\noindent
As the coefficient of the right-hand-side of eqn.~\ref{mpunto_de_val}
shows, the accretion rate and the mass gained or
lost by an IMBH of $10^2$--$10^4\,M_{\odot}$ during the past 10~Gyr is
negligible. Hence, we can conclude that the accretion of dark energy by an IMBH
does not affect the evolution of the accretor mass.\\
\section{Discussion and conclusions}
\label{conclusions}

In this work we investigated the spherical, steady-state accretion of
cosmological fluids onto an intermediate-mass black hole at the centre of a
globular cluster, taking into account the influence of the parent stellar
system on the accretion flow. We also include in our models a correction for
the effects of the motion of the black hole through the fluid. For
relativistics perfect fluids ($p \sim \rho$) and dark energy, our results show
that the presence of the cluster does not affect significantly the flow, hence
our results coincide with those derived from models of accretion onto isolated
black holes, such as those of  \citet{babichev}. The conclusions of these
authors hold, particularly that the accretion rate is proportional to
$(p_\infty + \rho_\infty) M^2$. As pointed out by \citet{babichev} this implies
that, depending on the nature of the dark energy, the IMBH mass can either
increase (for $\omega > -1$) or decrease (for $\omega < -1$). For $\omega =
-1$ the mass remains unchanged. For the particular case of IMBHs with masses in
the range $10^2$--$10^4\,M_{\odot}$ and a typical globular cluster, our
calculations show that the black hole mass is not affected by the accretion of
either dark energy or radiation from the cosmic microwave background.

However, for the accretion of cold dark matter the situation changes. We found
that if dark matter is collisionless or has a very low speed of sound, the
accretion rate no longer scales as the square of the black hole mass, but as
the square of the mass inside the critical radius. This is due to the effect of
the globular cluster gravitational field on the flow. As the critical radius
can be well outside the cluster core, this mass can be much greater than the
IMBH mass, and hence the accretion rate can be enhanced by a factor of the
order of $10^4$--$10^6$. We also found that the accretion rate scales as
$c_{\mathrm{s}}^{-3}$. These results lead to final IMBH masses greater than  $10^4\,
M_\odot$, well above the upper limits given by present observations. We point
out that this enhancement is independent of the details of our simple model,
particularly of the abrupt change in the critical radius observed as the
speed of sound decreases. Even if this change were replaced by a smooth
behaviour, the critical radius must approach the cluster outer regions as
$\omega$ decreases, greatly enhancing the accretion rate. Hence, there will
still be a lower limit for $\omega$ at which the mass accreted by the IMBH
during its lifetime exceeds the observational upper bounds. Our models estimate
this limit at $\omega \sim 10^{-9}$, which corresponds to a sound speed of
$c_{\mathrm s} \sim 10$~km/s. 
  
The last result has an important impact on our knowledge of IMBHs and dark
matter. If IMBHs exist at the centres of globular clusters and dark matter has
a low sound speed, IMBH
masses should have grown to values beyond the observed upper limits. To restore
the agreement with observations we are forced to assume that dark matter, if it
exists, must have a sound speed at least of the order of
$c_{\mathrm s} \sim 10$~km/s. Indeed, \citet{guzman} arrive at the same conclusion
investigating the accretion onto supermassive black holes using more
sophisticated, time dependent accreion models. Another possibility
to check our conclusions is to investigate the accretion onto stellar-mass
black holes. Some of these objects have known masses and their velocities are
well established. For example, \citet{mirabel} calculated the orbit of the
black hole X-ray nova XTE J1118+480. Computing the accretion rate for this
black hole (with a mass of $M_{\mathrm{BH}} = 6.9 M_{\odot}$ and a velocity of
$v_{\mathrm{BH}}\sim 140\mathrm{km/s}$, taken from these authors) we obtain
negligible accretion rates. Hence, stellar-mass black holes do not seem to be
useful for testing the dark matter equation of state. Thereby, if IMBHs were
finally detected at the centres of globular clusters and detailed models of
their accretion developed, these objects would turn into extraordinary tools to
investigate this issue.

\label{lastpage}

\end{document}